\documentstyle[12pt]{article} 
\newcommand{\beq}{\begin{equation}}
\newcommand{\eeq}{\end{equation}}
\newcommand{\beqa}{\begin{eqnarray}}
\newcommand{\eeqa}{\end{eqnarray}}
\newcommand{\ba}{\begin{array}}
\newcommand{\ea}{\end{array}}

\begin{document}

\begin{center}
{\large \bf Statistical Mechanics \\
of a Trapped Bose--Einstein Condensate}
\footnote{Meccanica Statistica 
di un Condensato di Bose--Einstein Intrappolato.} 
\footnote{Presented to SIMAI98, IV National Conference of the Societ\`a 
Italiana di Matematica Applicata ed Industriale, Giardini Naxos, 
Messina, June 1998.} 
\vskip 0.8 truecm 
{\bf Luca Salasnich} 
\vskip 0.5 truecm
Istituto Nazionale per la Fisica della Materia, Unit\`a di Milano,  \\
Dipartimento di Fisica, Universit\`a di Milano, \\
Via Celoria 16, 20133 Milano, Italy \\
Istituto Nazionale di Fisica Nucleare, Sezione di Padova, \\
Via Marzolo 8, 35131 Padova, Italy \\
Dipartimento di Matematica Pura ed Applicata, \\
Universit\`a di Padova, Via Belzoni 7, 35131 Padova, Italy 
\end{center}

\vskip 1. truecm

\par
Bose--Einstein condensation (BEC) in a gas has now been achieved [1]. 
Alkali atoms ($^{87}Rb$, $^{23}Na$ and $^{7}Li$) have been cooled to the 
point of condensation (temperature of $100$ nK) using laser cooling 
and trapping, followed by magnetic trapping and evaporative cooling. 
This important experimental result has also renewed the interest 
on theoretical studies of BEC [2]. 
In this contribution we discuss the statistical mechanics 
of the trapped BEC at zero temperature [3]. 
In particular, we study the stability of 
the condensate by using a variational method with 
local and non--local interaction between the particles [4]. 
\par
The quantum Hamiltonian operator of a gas of $N$ identical bosonic particles 
of mass $m$ and without spin in a trap is given by 
\beq
{\hat H}=\sum_{i=1}^N \Big( -{\hbar^2\over 2 m} \nabla_i^2 
+ V_0({\bf r}_i) \Big) + 
{1\over 2} \sum_{ij=1}^N V({\bf r}_i,{\bf r}_j) \; ,
\eeq
where $V_0({\bf r})$ is the trapping potential and $V({\bf r},{\bf r}')$  
is the interaction potential [3]. 
Let $\Psi_{\alpha} ({\bf r})$ be the single particle wave function, 
where $\alpha$ represents a complete set of quantum numbers. 
The totally symmetric many--particle wave function with quantum numbers 
$\alpha_1,...,\alpha_N$ results 
\beq 
\Phi_{\alpha_1 ... \alpha_N}({\bf r}_1,...,{\bf r}_N) = 
\sqrt{N_{\alpha_1}!...N_{\alpha_N}!\over N!} 
\sum_{{\hat P}} \Psi_{{\hat P}\alpha_1}({\bf r}_1)...
\Psi_{{\hat P}\alpha_N}({\bf r}_N) \; ,
\eeq
where ${\hat P}$ is the permutation operator 
and $N_{\alpha_i}$ is the number of particles in the state $\alpha_i$.  
The system is a Bose--Einstein condensate 
if all the $N$ particles are in the same 
quantum state $\alpha$. Thus, for $\alpha =\alpha_1$, we have  
$N_{\alpha_1}=N$, $N_{\alpha_2}=...=N_{\alpha_N}=0$ and 
the many--body wave function reads 
\beq 
\Phi_{\alpha}({\bf r}_1,...,{\bf r}_N) = 
\Psi_{\alpha}({\bf r}_1)...\Psi_{\alpha}({\bf r}_N) \; . 
\eeq
We observe that the mean energy per particle, which is given by
\beq
\epsilon =<\Phi_{\alpha} |{{\hat H}\over N}|\Phi_{\alpha} > = {1\over N} 
\int d^3{\bf r}_1 ... d^3{\bf r}_N \Phi_{\alpha}^*({\bf r}_1,...,{\bf r}_N) 
{\hat H} \Phi_{\alpha}({\bf r}_1,...,{\bf r}_N) \; ,
\eeq
can be written as
\beq 
\epsilon [\Psi_{\alpha} ]
= \int d^3{\bf r} \; {\hbar^2\over 2m} |\nabla \Psi_{\alpha} 
({\bf r})|^2 
+ V_0({\bf r}) |\Psi_{\alpha} ({\bf r})|^2 
+ {(N-1)\over 2} 
\int d^3{\bf r} d^3{\bf r}' |\Psi_{\alpha} ({\bf r})|^2 V({\bf r},{\bf r}')
|\Psi_{\alpha} ({\bf r}')|^2 \; , 
\eeq
where $\epsilon$ is a functional of the wave function $\Psi_{\alpha}$. 
By considering a weakly interacting gas, such that 
$V({\bf r},{\bf r}') = B \delta^3 ({\bf r}-{\bf r}')$, 
where $B={4\pi \hbar^2 a_s/m}$ is the scattering amplitude, 
the mean energy per particle results 
\beq
\epsilon [\Psi ] = \int d^3{\bf r} \; {\hbar^2\over 2m} 
|\nabla \Psi ({\bf r})|^2 
+ V_0({\bf r}) |\Psi ({\bf r})|^2 +{BN\over 2} |\Psi ({\bf r})|^4 \; ,
\eeq 
where we have omitted the quantum number $\alpha$ and we have written 
$N$ instead of $N-1$ because we consider a large number of particles. 
This is the so--called Gross--Pitaevskii functional [5]. 
We observe that the scattering length $a_s$ is supposed to be positive for 
$^{87}Rb$ and $^{23}Na$, but negative for $^{7}Li$. 
It means that for $^{87}Rb$ and $^{23}Na$ the interatomic 
interaction is repulsive while for $^{7}Li$ the atom--atom 
interaction is effectively attractive [1]. 
\par
To find the minimum of the energy functional $\epsilon [\Psi ]$ 
we impose that the first order variation of the energy functional 
is zero 
\beq
\delta_{\Psi} \epsilon [\Psi ] = 0 
\eeq
with the constraint
\beq
\int d^3{\bf r} |\Psi ({\bf r})|^2 = 1 \; .
\eeq
In this way we get the following Euler--Lagrange equation
\beq
\Big[ -{\hbar^2\over 2m} \nabla^2 
+ V_0({\bf r}) + B N |\Psi ({\bf r})|^2 \Big] \Psi ({\bf r}) 
= \mu \Psi ({\bf r}) \; ,
\eeq
where $\mu$ (Lagrange multiplier) is the chemical potential. 
This equation (Gross--Pitaevskii equation) 
has the form of a nonlinear stationary Schr\"odinger equation [5]. 
\par
We analyze the Gross--Pitaevskii functional by using a variational technique. 
The external potential of the trap is well approximated by 
an harmonic oscillator $V_0({\bf r})=(m\omega^2/2)r^2$, where 
the frequency $\omega$ is proportional to the amplitude of the trapping 
magnetic field. 
We choose a Gaussian trial wave function for the BEC with a free 
parameter $\sigma$, which is the standard deviation of the Gaussian, i.e. 
the mean radius of the condensate
\beq
\Psi ({\bf r}) = {1\over \pi^{3/4}\sigma^{3/2} } 
\exp{({-r^2\over 2\sigma^2})} \; .
\eeq
The resulting energy function is 
\beq
\epsilon (\sigma )={3\over 2}({\hbar^2 \over 2m}) {1\over \sigma^2} 
+{3\over 2}({m\omega^2\over 2}) \sigma^2 + 
{BN \over 2} {1\over (2\pi )^{3/2} \sigma^3} \; .
\eeq
We study when the first derivative of $\epsilon (\sigma )$ 
is zero and we obtain the formula 
\beq
N= {2 (2\pi )^{3/2}\over B}[ ({m\omega^2\over 2}) \sigma^5 - 
({\hbar^2 \over 2 m})\sigma] \; . 
\eeq
For $B>0$ there is only one solution, which is 
a minimum of the energy functional (stable solution), 
instead for $B<0$ there 
are two solutions: one is a maximum (unstable solution) 
and the other a minimum (metastable solution) of the energy functional. 
In fact, for $B<0$ the absolute minimum of the energy function is 
$\epsilon = \infty$ at $\sigma =0$. 
It follows that for $B>0$, when $N=0$ we have $\sigma = 
\sqrt{\hbar/( m\omega )}$ and $\sigma \to \infty$ for $N \to \infty$. 
For $B<0$ the mean radius $\sigma$ decreases by increasing 
the number of bosons $N$ to a minimum radius $\sigma^{min}$, 
with a maximum number $N^{max}$ of bosons. For greater values of $N$ 
the condensate becomes unstable. 
It is not difficult 
to obtain this critical number of bosons. We put ${\tilde B}=-B$ and get 
\beq
0={dN \over d\sigma}={2 (2\pi )^{3/2}\over {\tilde B}}
[({\hbar^2 \over 2 m}) - 5({m\omega^2\over 2}) \sigma^4] \; ,
\eeq
from which we have the minimum radius 
$\sigma^{min}=5^{-1/4} \sqrt{\hbar /(m\omega )}$ 
and the maximum number of bosons 
\beq
N^{max} = {8 \over 5^{5/4}} {(2\pi )^{3/2}\over {\tilde B}} 
{\hbar^2\over 2 m} \sqrt{\hbar \over m\omega } \; .
\eeq 
Thus we obtain an analytical formula of the maximum number 
$N^{max}$ of bosons 
for which the condensate, with attractive interaction, is metastable. 
\par
We observe that in general we can not neglect the non--zero range 
dependence of the interaction potential. Let us now consider a more realistic 
non--local interaction between the particles 
\beq
V({\bf r},{\bf r}')= B \delta^3 ({\bf r}-{\bf r}') - 
A {\exp{(-\Gamma |{\bf r}-{\bf r}'|)} \over |{\bf r}-{\bf r}'| } \; ,
\eeq 
where the constants $A$, $B$ and $\Gamma$ are chosen to reproduce 
phenomenologically the experimental data [4]. 
By using again the trial Gaussian wave function 
and studying when the first 
derivative of the energy function $\epsilon (\sigma )$ 
is zero we get the following formula
\beq
N = { \big({m\omega^2\over 2}\big)\sigma^4-\big({\hbar^2\over 2 m} \big)
\over {B\over 2(2\pi )^{3/2}}\sigma^{-1} + 
{1\over 3} {A \Gamma^2\over \sqrt{2\pi}} 
\sigma^3 - {A\over 6}\sqrt{2\over \pi}\sigma - {A\Gamma^3\over 6}\sigma^4 
\exp{({\sigma^2 \Gamma^2\over 2})} erfc({\sigma \Gamma  \sqrt{2}}) }
\; ,
\eeq
where $erfc(x)=1-erf(x)$ is the complementary error function. 
For $B>0$ there is only one solution, which is 
a minimum (stable solution), instead for $B<0$ there are three solutions: 
one is a maximum (unstable solution) 
and the other two are minima (metastable solutions). 
Thus, we find that with a non--local interaction 
for $B<0$ there is a new metastable branch at high density.  
\par
To summarize, we study the stability of a gas of 
weakly interacting bosons in a harmonic trap with local and non--local 
interaction. To minimize 
the energy functional, we choose a Gaussian trial wave function, 
where its standard deviation represents the mean radius of the condensate. 
In the case of repulsive interaction the mean radius of the 
condensate grows by increasing the number of bosons. 
In the case of attractive interaction 
the mean radius decreases by increasing the number of bosons until 
it becomes unstable. 
\par
Our variational technique can be applied also to systems with 
different shapes of the trapping potential. 
In the future it will be interesting to investigate the time evolution 
of the condensate with exotic initial conditions. 

\vskip 0.5 truecm
\par
This work has been supported by INFM under the Research Advanced Project (PRA) 
on "Bose--Einstein Condensation". The author is grateful 
to A. Parola and L. Reatto for stimulating discussions and 
thanks V.R. Manfredi and G. Maino for their interest on the subject 
of this contribution. 

\vskip 0.5 truecm

\begin{description}

\item{\ [1]} M.H. Anderson, et al., Science {\bf 269}, 189 (1995); 
K.B. Davis, et al., Phys. Rev. Lett. {\bf 75}, 
3969 (1995); C.C. Bradley, et al., Phys. Rev. Lett. {\bf 75}, 1687 (1995). 

\item{\ [2]} F. Dalfovo and S. Stringari, Phys. Rev. A {\bf 53}, 2477 (1996); 
C.C. Bradley, C.A. Sackett, and R.G. Hulet, Phys. Rev. Lett. {\bf 78}, 
985 (1997). 

\item{\ [3]} A.L. Fetter and J.D. Walecka, {\it Quantum Theory 
of Many--Particle Systems} (McGraw--Hill, New York, 1971); 

\item{\ [4]} A. Parola, L. Salasnich and L. Reatto, to appear in Phys. Rev. A. 

\item{\ [5]} E.P. Gross, Nuovo Cimento {\bf 20}, 454 (1961); 
L.P. Pitaevskii, Sov. Phys. JETP {\bf 13}, 451 (1961). 

\end{description}

\end{document}